\documentclass[sigconf]{acmart}
\copyrightyear{2025}
\acmYear{2025}
\setcopyright{rightsretained}
\acmConference[CHI EA '25]{Extended Abstracts of the CHI Conference on Human Factors in Computing Systems}{April 26-May 1, 2025}{Yokohama, Japan}
\acmBooktitle{Extended Abstracts of the CHI Conference on Human Factors in Computing Systems (CHI EA '25), April 26-May 1, 2025, Yokohama, Japan}\acmDOI{10.1145/3706599.3721343}
\acmISBN{979-8-4007-1395-8/2025/04}

\begin{document}

\newcommand{\cl}{Clo(o)k}
\title{\cl{}: Human-Time Interactions Through a Clock That ``Looks''}

\author{Zhuoyue Lyu}
\authornote{This project was completed while the author was a student at MIT.}
\affiliation{%
 \institution{University of Cambridge}
 \city{Cambridge}
 \country{United Kingdom}}
  \email{zl536@cam.ac.uk}

\begin{teaserfigure}
    \centering
    \includegraphics[height=0.17704\textwidth]{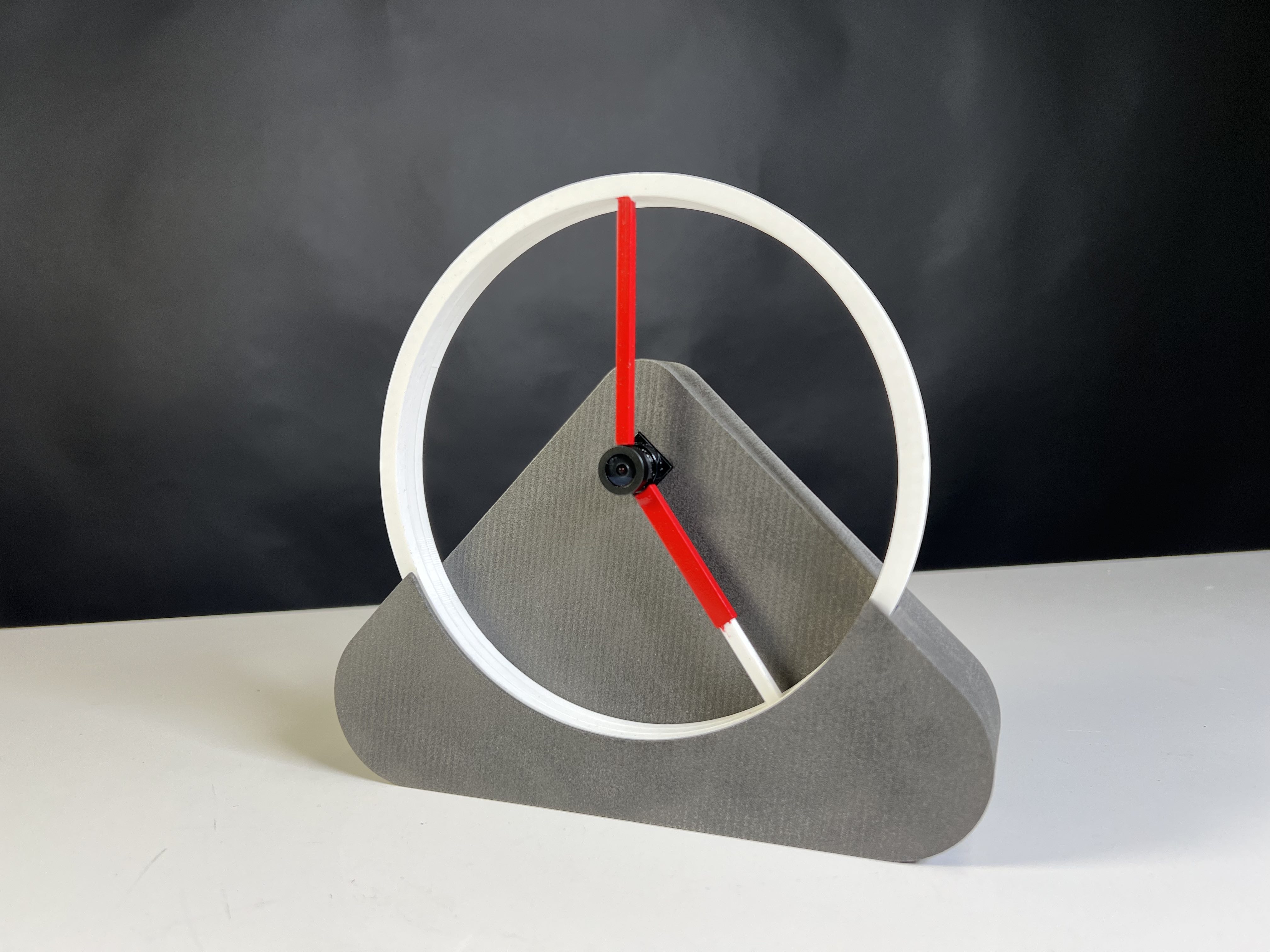} \hfill
    \includegraphics[height=0.17704\textwidth]{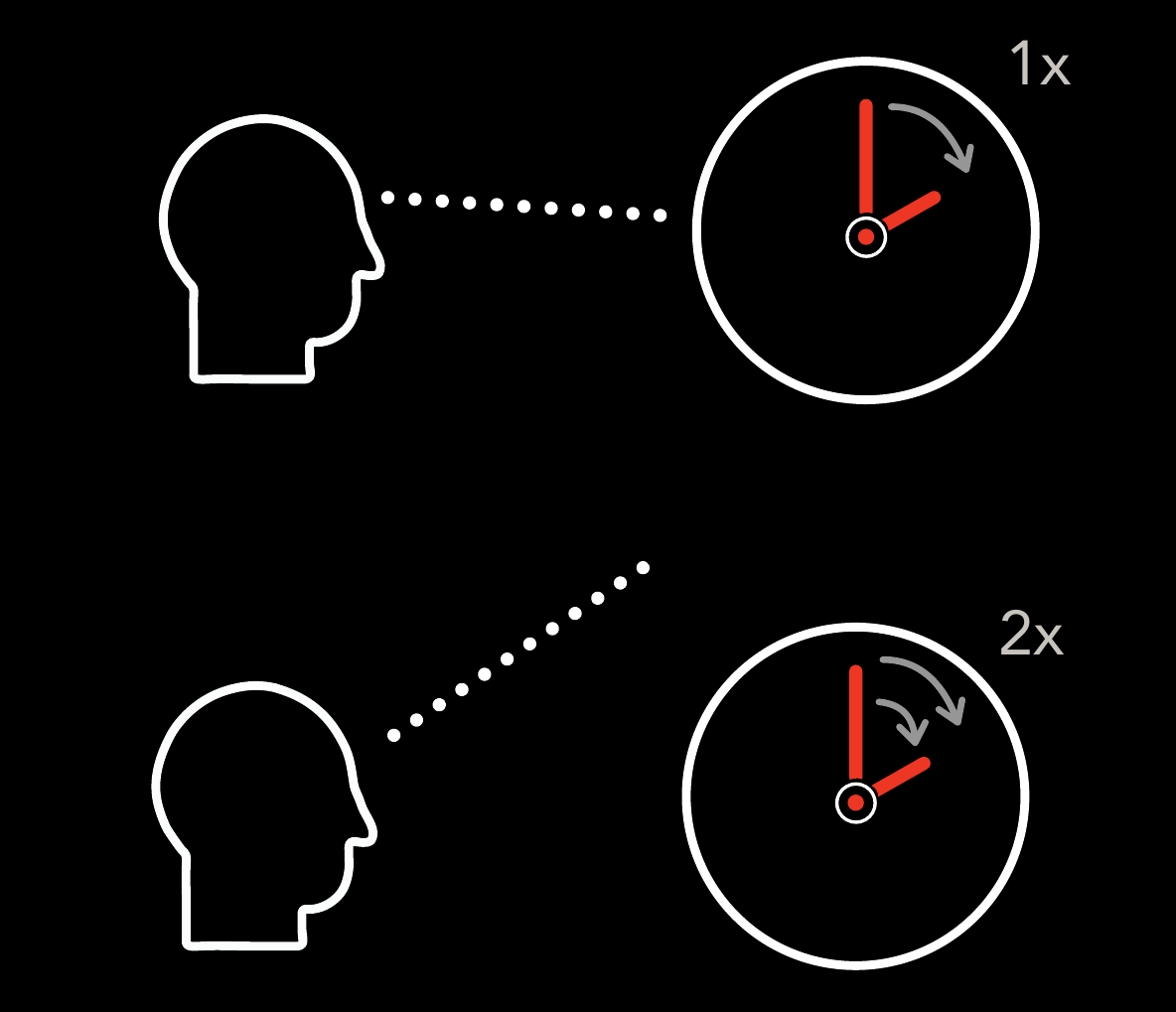} \hfill
    \includegraphics[height=0.17704\textwidth]{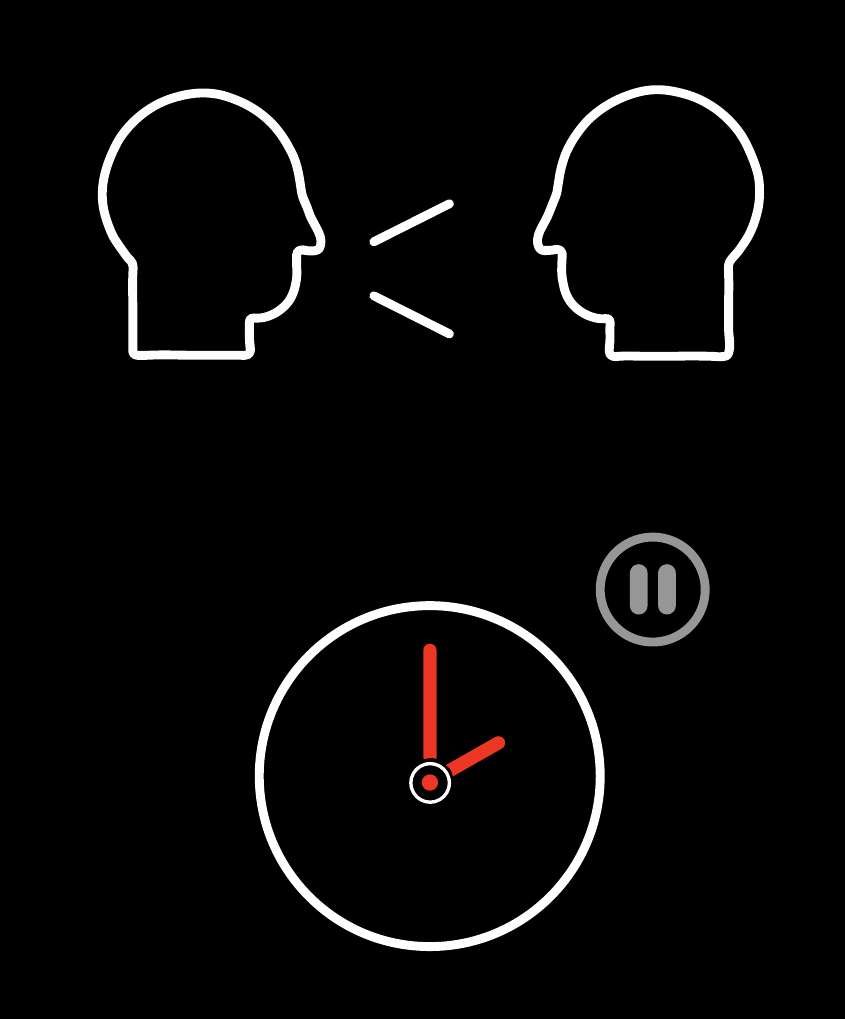} \hfill
    \includegraphics[height=0.17704\textwidth]{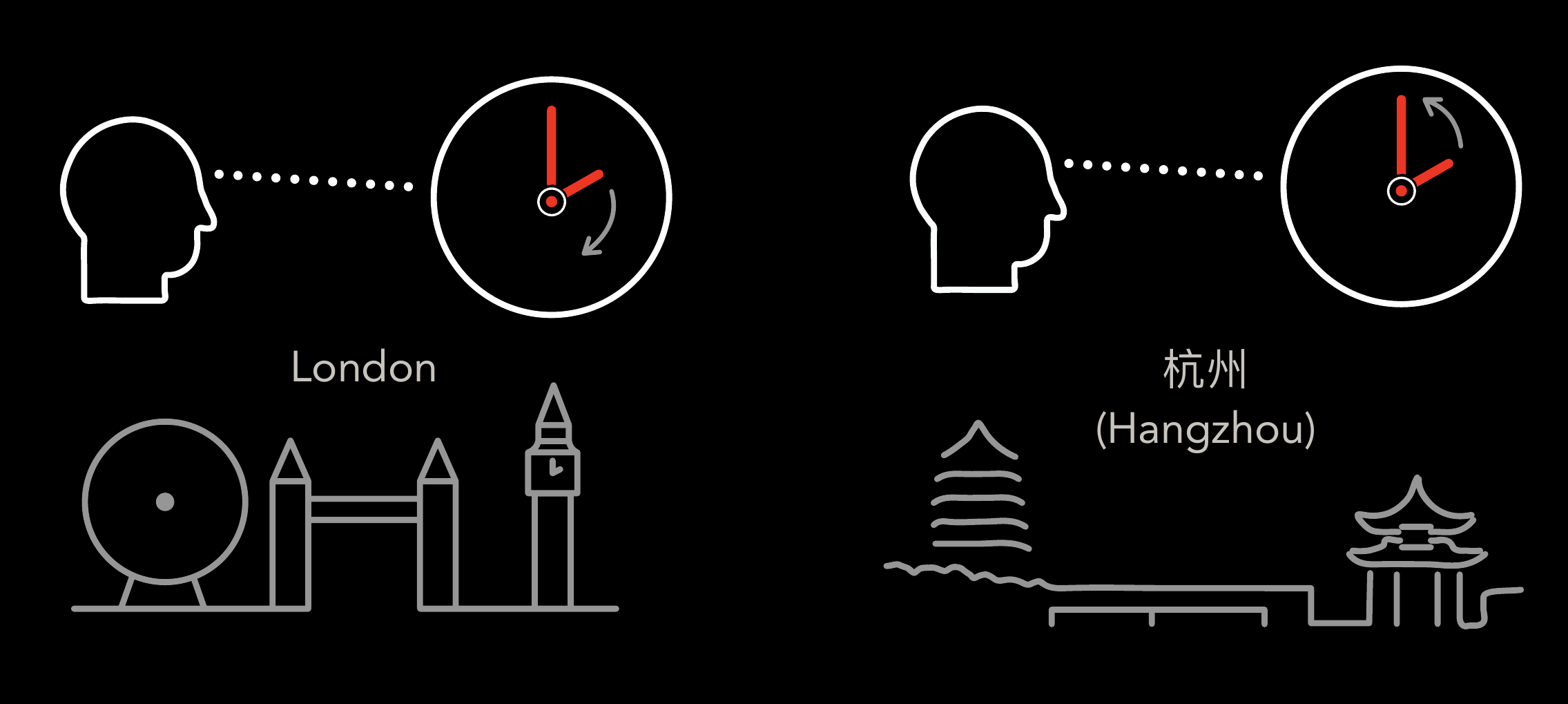}
    \caption{Clo(o)k and its three interaction types (Section~\ref{sec:interactions}). Video demo, design files, and code available at \href{https://zhuoyuelyu.com/clook}{\textbf{zhuoyuelyu.com/clook}}}
    \Description{This image presents Clo(o)k, a uniquely designed interactive clock with a triangular base and a circular frame, blending functionality with mindful engagement. The first subfigure displays the physical clock, featuring a minimalist design with red clock hands, a white circular rim, and a gray base. The second subfigure illustrates the first type of interaction: Clo(o)k operates like a normal clock when the user looks at it but speeds up when their attention shifts away. The third subfigure demonstrates its second type of interaction—if it detects the user engaged in a conversation, it pauses to alleviate time pressure and encourage uninterrupted communication. The final subfigure highlights its third type of interaction: when two users in different locations (e.g., London and Hangzhou) simultaneously look at their respective Clo(o)ks, they will see each other’s current time, fostering a shared temporal connection despite the physical distance.}
    \label{fig:interactions}
\end{teaserfigure}


\begin{abstract}
What if a clock could do more than tell time—what if it could look around? This project explores the conceptualization, design, and construction of a timepiece with visual perception capabilities, featuring three types of human-time interactions. Informal observations during a demonstration highlight its unique user experiences.
\end{abstract}

\ccsdesc[500]{Human-centered computing~Interaction devices}
\keywords{clock, time, tangible, telepresence}

\maketitle

\section{Introduction and Related Work}
Time is a multifaceted concept, comprising both objective and subjective elements. Objectively, it is precisely measured by the ticking of clocks, while subjectively it can slip by unnoticed when we are deeply engrossed in our activities. On a global scale, time unites us: when it is 3 AM in Boston, it is simultaneously 3 PM in Beijing~\footnote{During U.S. daylight saving time (March–November).}—an understanding that enables seamless communication across time zones. In popular culture, Disney films such as \textit{Snow White and the Seven Dwarfs}\footnote{\url{https://movies.disney.com/snow-white-and-the-seven-dwarfs}} and \textit{Beauty and the Beast}\footnote{\url{https://movies.disney.com/beauty-and-the-beast}} imagine everyday objects with human personalities. Inspired by these portrayals, we wondered what would happen if a clock could see and interact with people. This notion opens creative possibilities for playful, human-clock interactions~\cite{mcgrath1986time}, motivating the design of the Clo(o)k—a clock that looks. 

Previous research has explored practical uses of everyday objects~\cite{10.1145/3544548.3581449,lyu2025objesturesbimanualinteractionseveryday,10.1145/3544549.3585871}. Specifically, clocks have been employed to visualize upcoming events~\cite{10.1145/506443.506505}, track sleep patterns~\cite{10.1145/1517664.1517672}, and support planning and reflection~\cite{10.1145/3334480.3382830,10.1145/3357236.3395439}. This work takes a more playful approach by creating an interactive experience that allows users to reflect on the concept of time. Telepresence technology has used various sensory modalities, including sound~\cite{10.1145/1935701.1935705}, touch~\cite{10.1145/1120212.1120435,10.1145/2642918.2647377}, and sight~\cite{10.1145/142750.142977} to enable people to feel the presence of others even when they are far apart. Building on this, we integrate time into interactions, enabling new possibilities across time and space.



\section{\cl{} Interactions}\label{sec:interactions}
The Clo(o)k operates like a normal clock when the user is looking at it, but speeds up when the user's attention shifts away (2nd subfigure in Figure~\ref{fig:interactions}). If the Clo(o)k notices that the user is engaged in conversation with someone, it stops moving to alleviate any concerns about time and allow for uninterrupted communication (3rd subfigure in Figure~\ref{fig:interactions}). If the user's loved one resides in a different time zone (e.g., London, UK, and Hangzhou, China), when both parties look at their respective Clo(o)ks simultaneously, they will see each other's current local time (4th subfigure in Figure~\ref{fig:interactions}).

\begin{figure*}
    \centering
    \begin{minipage}{\textwidth}
        \centering
        \includegraphics[height=0.1475\textwidth]{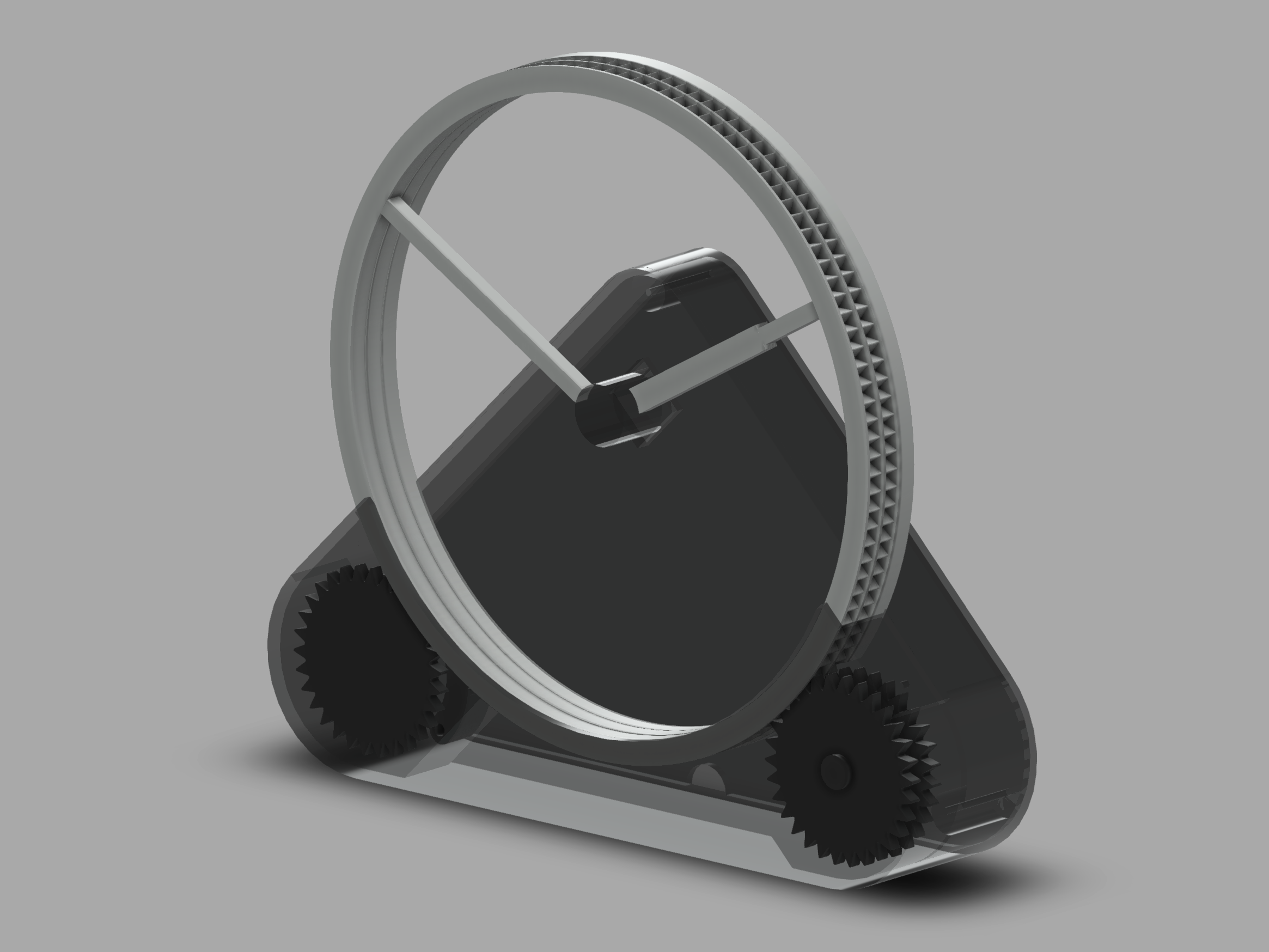} \hfill
        \includegraphics[height=0.1475\textwidth]{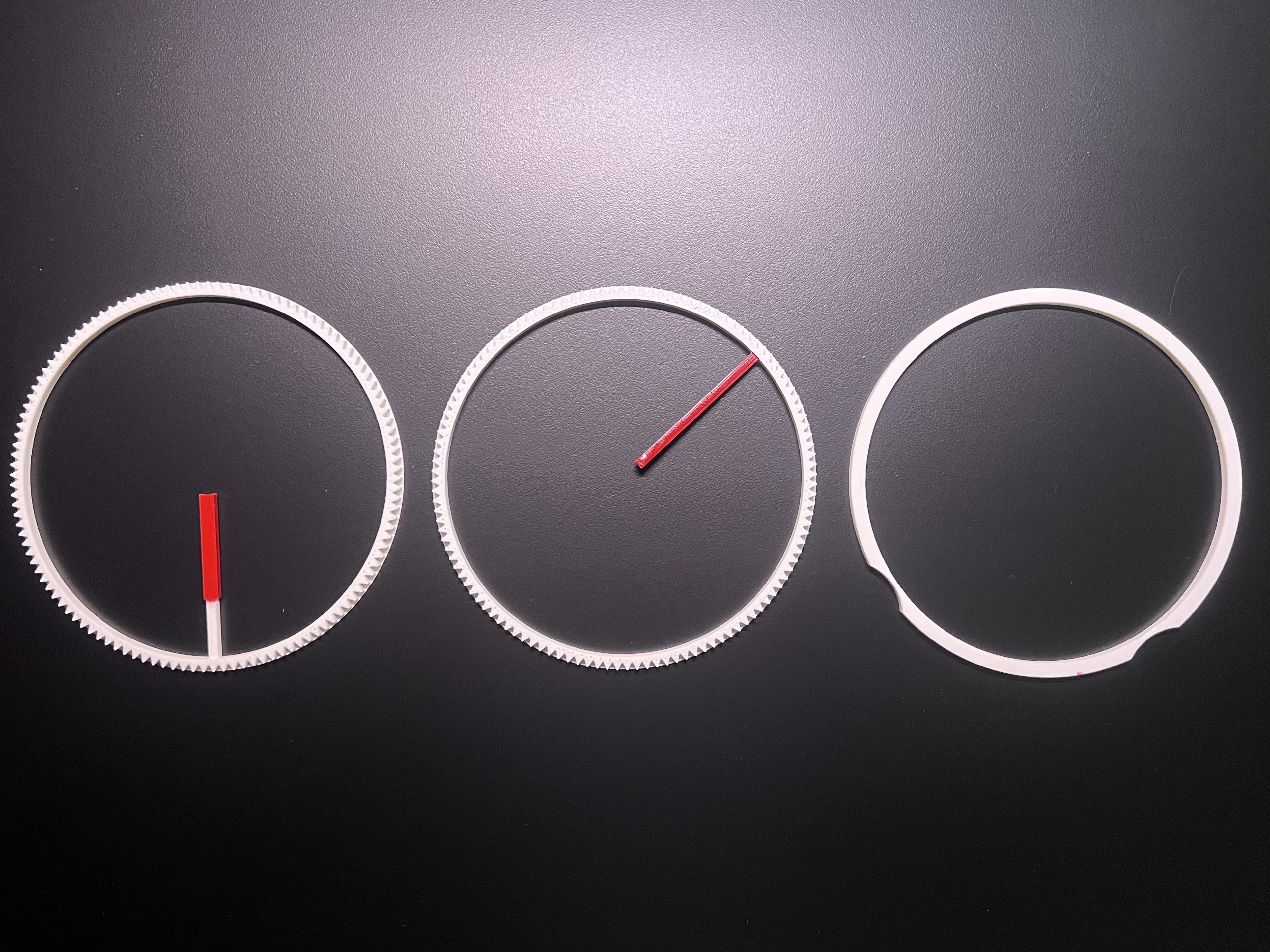} \hfill
        \includegraphics[height=0.1475\textwidth]{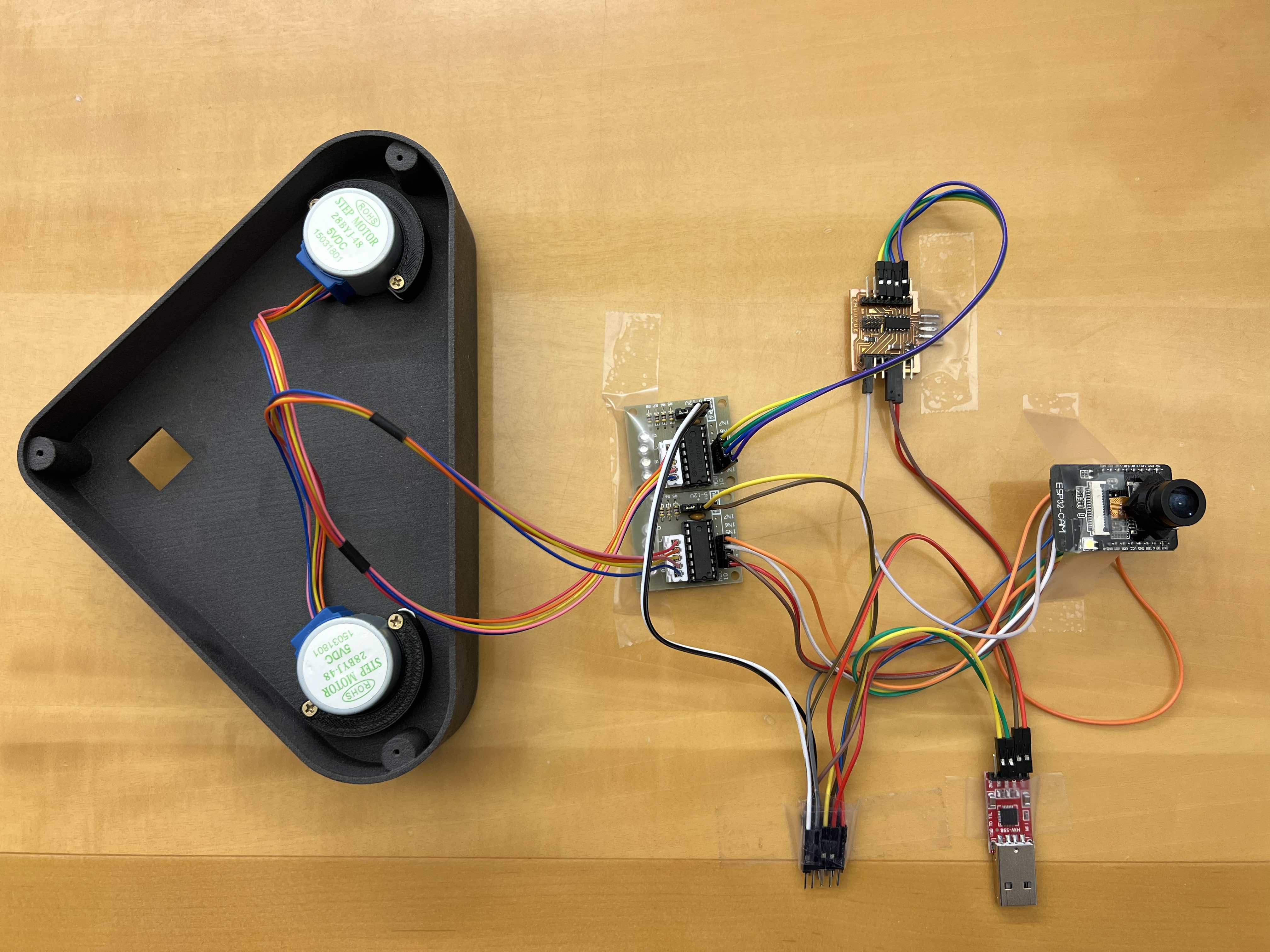} \hfill
        \includegraphics[height=0.1475\textwidth]{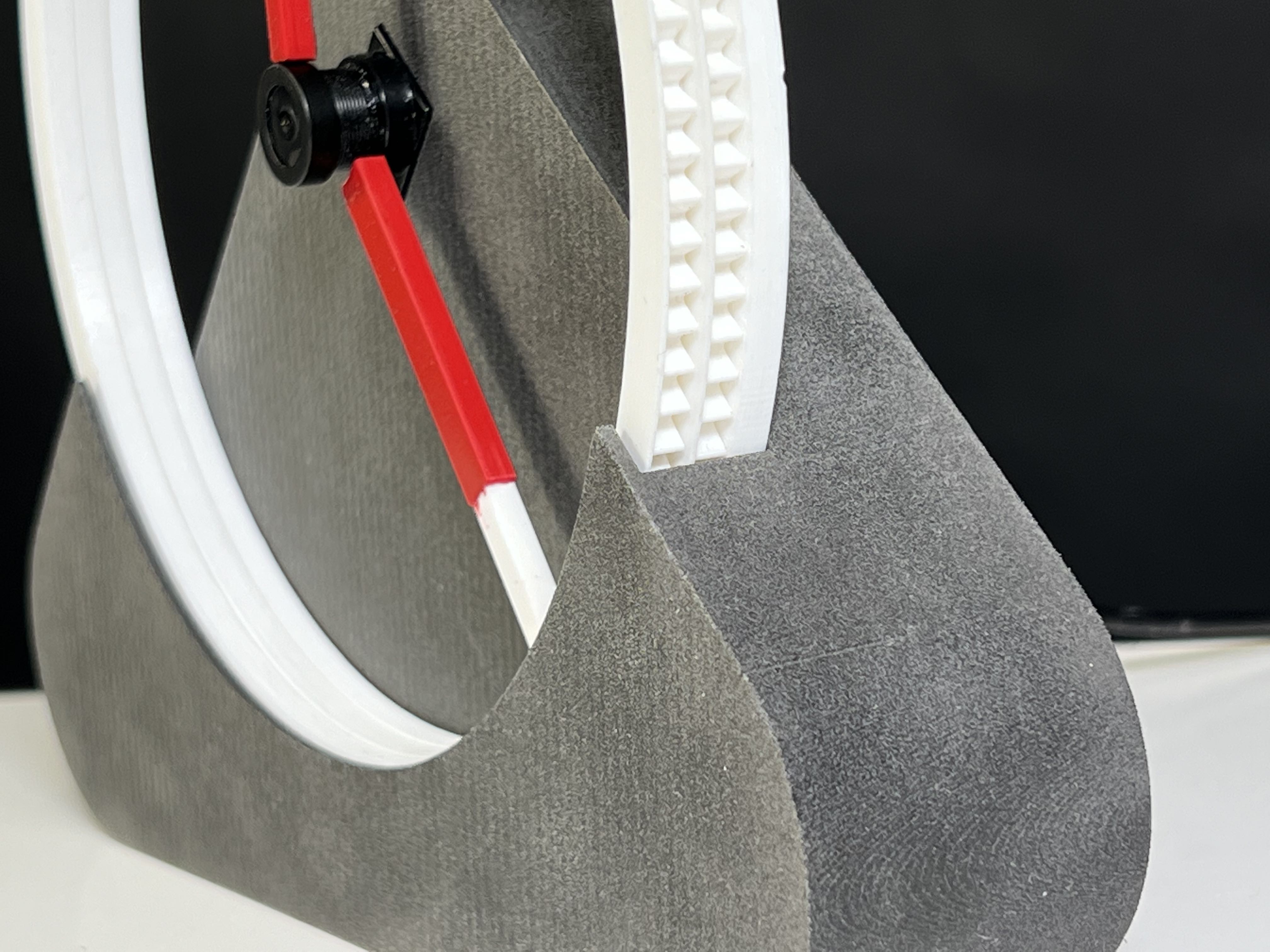} \hfill
        \includegraphics[height=0.1475\textwidth]{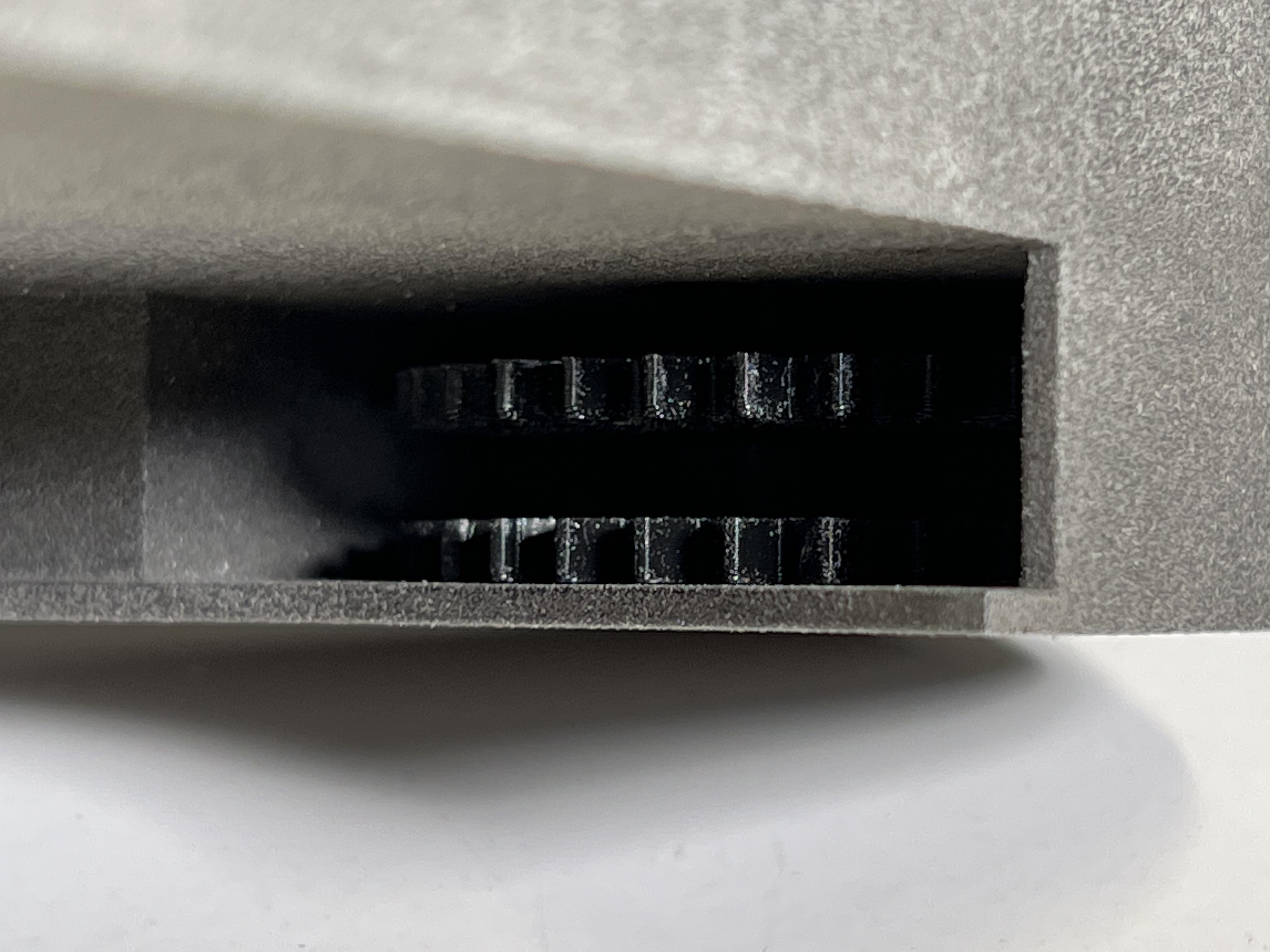} 
    \end{minipage}

    \vspace{0.4mm} 

    \begin{minipage}{\textwidth}
        \centering
        \includegraphics[height=0.1475\textwidth]{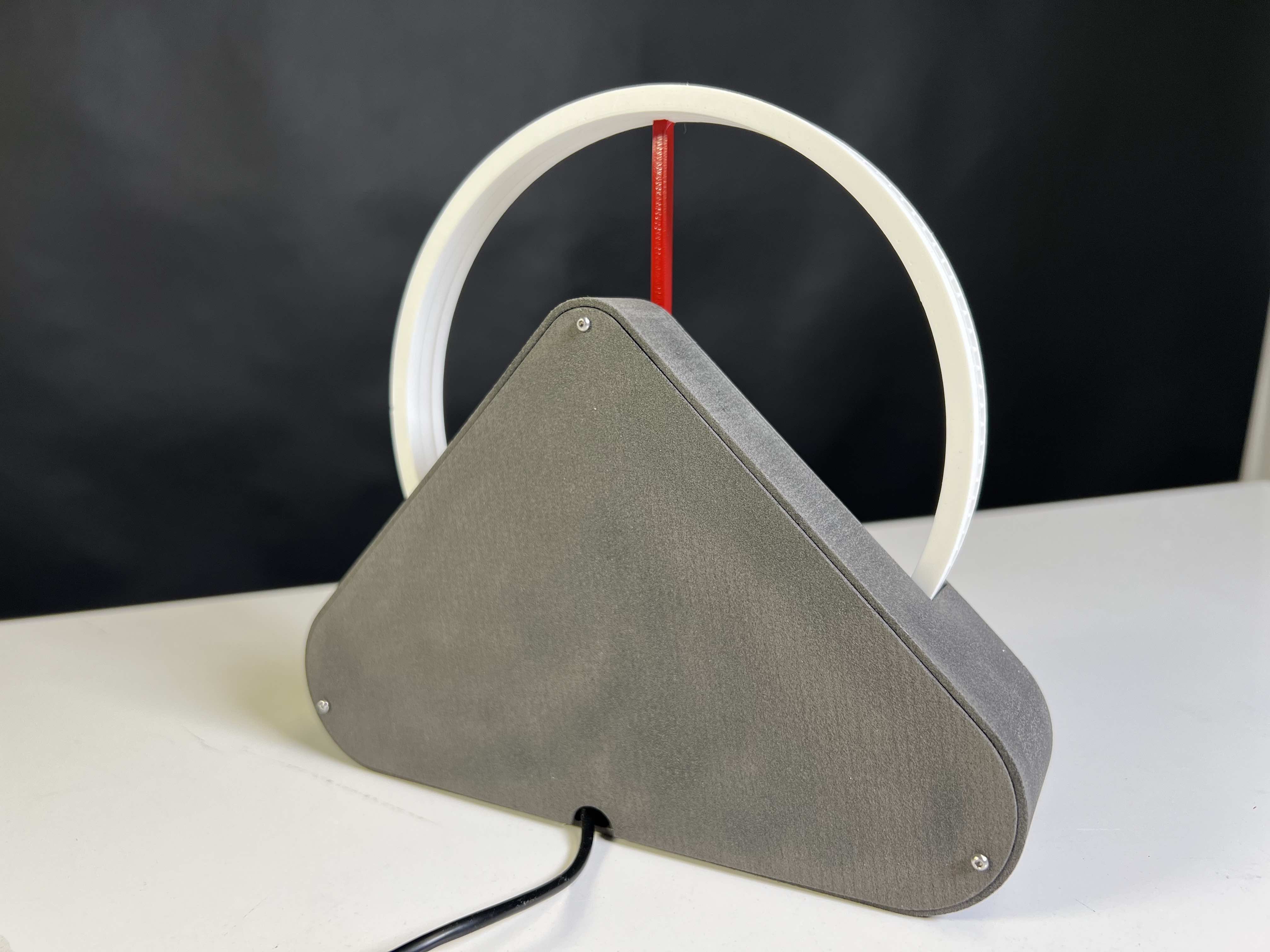} \hfill
        \includegraphics[height=0.1475\textwidth]{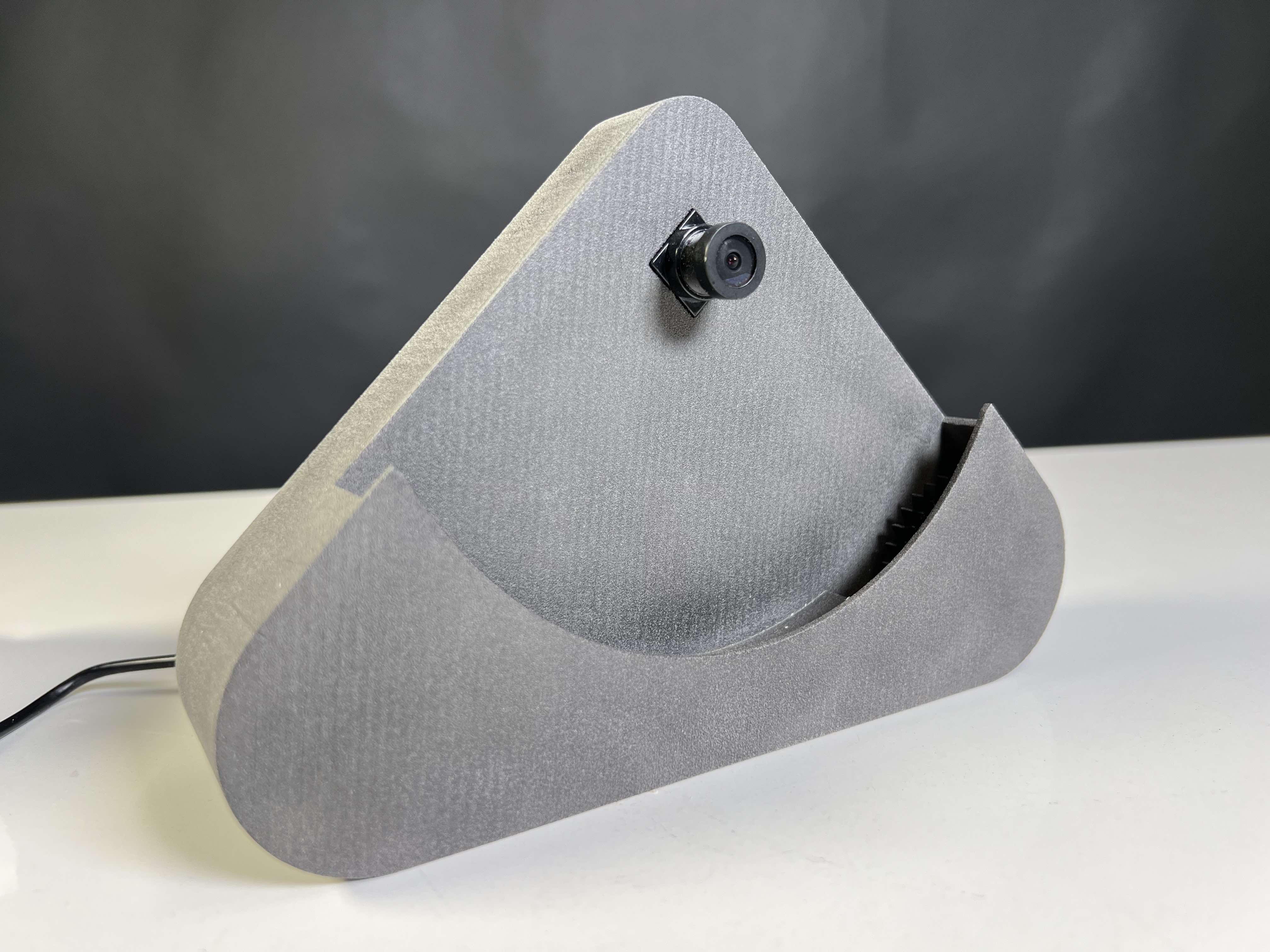} \hfill
        \includegraphics[height=0.1475\textwidth]{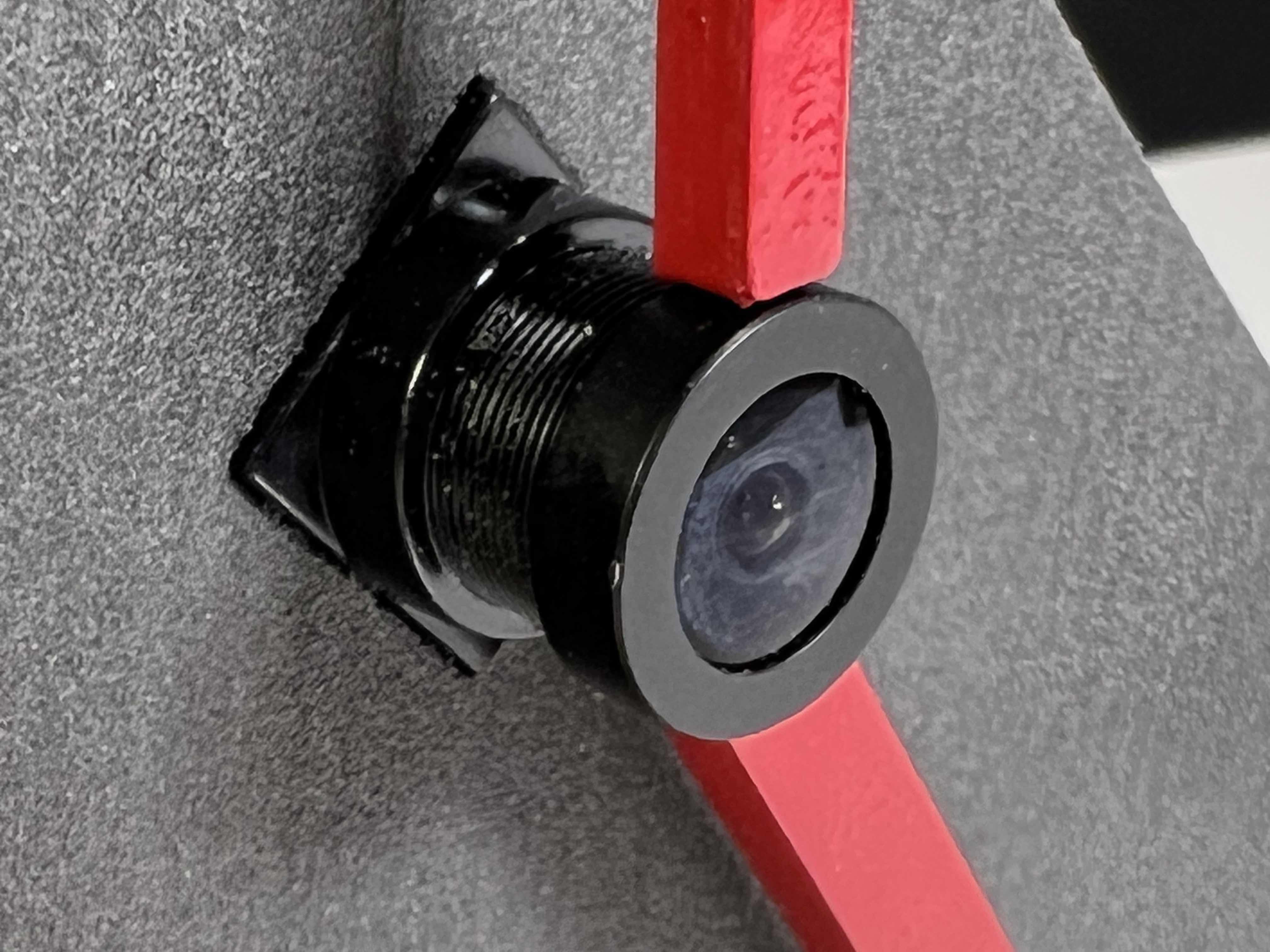} \hfill
        \includegraphics[height=0.1475\textwidth]{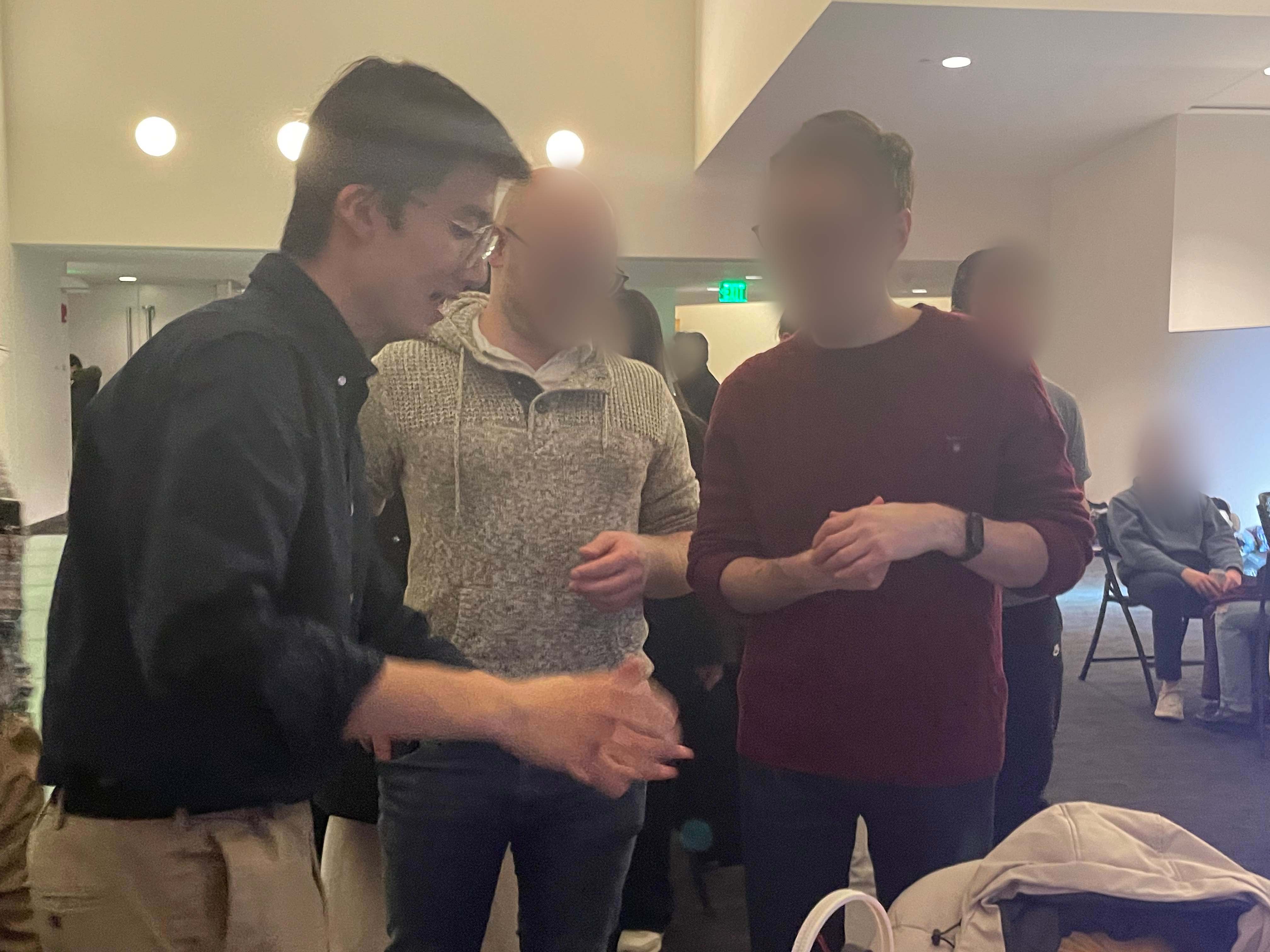} \hfill
        \includegraphics[height=0.1475\textwidth]{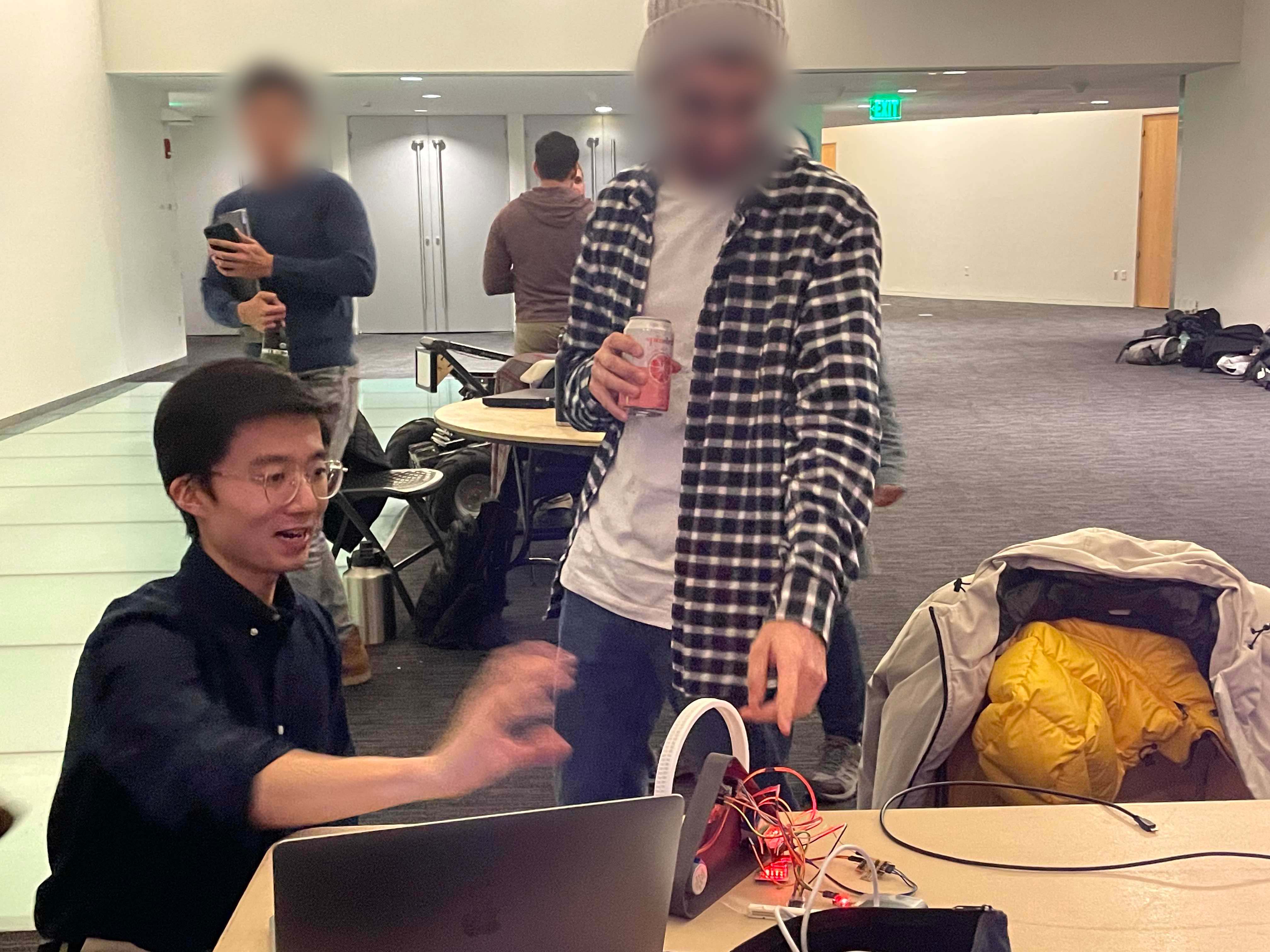}
    \end{minipage}
    \caption{Design and fabrication of the Clo(o)k, along with its demo at the MAS.863 Open House (Dec. 20, 2022, MIT Media Lab).}
    \Description{This image series documents the design, fabrication, and demonstration of Clo(o)k. The first row features a 3D-rendered model, internal electronic components, a close-up of the structure, and the gear system housed within the base. The second row showcases the physical prototype from multiple angles, emphasizing its triangular form, integrated camera, and material texture. The final images capture the author presenting Clo(o)k at the MAS.863 Open House, engaging with attendees and explaining its functionality, highlighting its interactive and temporal features in a live demonstration setting.}
    \label{fig:details}
\end{figure*}

\section{Design and Fabrication}
To achieve a simple, clean design for \cl{}, basic shapes such as circles and triangles were chosen for the clock face and base. Positioning the camera at the center preserves the appearance of a standard clock, allowing users to initially perceive it as an ordinary timepiece. The twist emerges when they realize the central element is a camera. However, this placement precluded the use of a traditional clock mechanism, where all hands attach at the center. Instead, the hollow clock mechanism~\footnote{\url{https://www.thingiverse.com/thing:4761858}} was adopted, with the minute and hour hands integrated into rings driven by concealed gear sets in the base (Figure~\ref{fig:details}). The 3D design was created using Fusion 360~\footnote{\url{https://www.autodesk.com/products/fusion-360/}} and Blender~\footnote{\url{https://www.blender.org/}}, and then exported as STL files to be printed. The rings and gears were printed with PLA using Prusa~\footnote{\url{https://www.prusa3d.com/}}, while the hands were spray-painted red. The base was printed using Fuse 1~\footnote{\url{https://formlabs.com/3d-printers/fuse-1/}} with powder for better quality.

\section{Hardware and Software}
\cl{} utilizes an ESP32-CAM~\footnote{\url{https://docs.ai-thinker.com/en/esp32-cam}} with a modified lens to capture the video, control one of the stepper motors, and wirelessly communicate with the computer and other Clo(o)ks. A D11C~\footnote{\url{https://ww1.microchip.com/downloads/en/DeviceDoc/Atmel-42363-SAM-D11_Summary.pdf}} board was designed and milled to extend the pins of the ESP32-CAM, enabling it to receive serial signals from the ESP32-CAM and control another stepper motor. Two 28BYJ-48 stepper motors~\footnote{\url{https://components101.com/motors/28byj-48-stepper-motor}} with driver boards control the rings. An FTDI Converter~\footnote{\url{https://microcontrollerslab.com/ftdi-usb-to-serial-converter-cable-use-linux-windows/}} programs the ESP32-CAM and supplies power (Figure~\ref{fig:details}). Both ESP32-CAM and D11C were programmed using Arduino~\footnote{\url{https://www.arduino.cc/}}. Although there is a reduced version of OpenCV~\footnote{\url{https://github.com/joachimBurket/esp32-opencv}} for ESP32-CAM, for more flexibility and convenience,  a separate laptop running Python script using OpenCV~\footnote{\url{https://opencv.org/}} is used for facial detection, which receives the video stream from ESP32-CAM wirelessly. ESP32-CAM updates the movements of the stepper motors based on the information (number of faces) it receives from Python through the USB serial connection.

\section{Informal Observations and Limitations}
\cl{} was demonstrated during an Open House at the MIT Media Lab (Figure~\ref{fig:details}). Attendees consistently responded positively upon learning about the Clo(o)k's interactions, particularly appreciating the exploration of time’s dual nature (subjective and objective). Many attendees were drawn to the minimalist design and expressed pleasant surprise upon discovering the detachable ring mechanism. Several international students with loved ones in different time zones highlighted the usefulness of the third interaction, noting its ability to evoke a sense of connection with distant family and friends. However, some limitations were identified: low lighting hindered facial recognition accuracy, and latency in serial communication between the ESP32-CAM and D11C caused slight delays in the minute and hour hands' movements.

\section{Conclusion and Future Work}
\begin{figure}[!htb]
    \centering
    \begin{minipage}{\linewidth}
        \centering
        \includegraphics[height=0.212\textwidth]{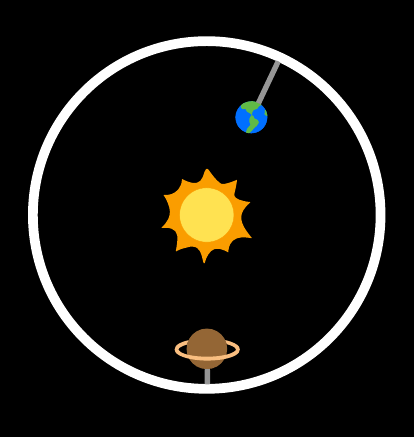} \hfill
        \includegraphics[height=0.212\textwidth]{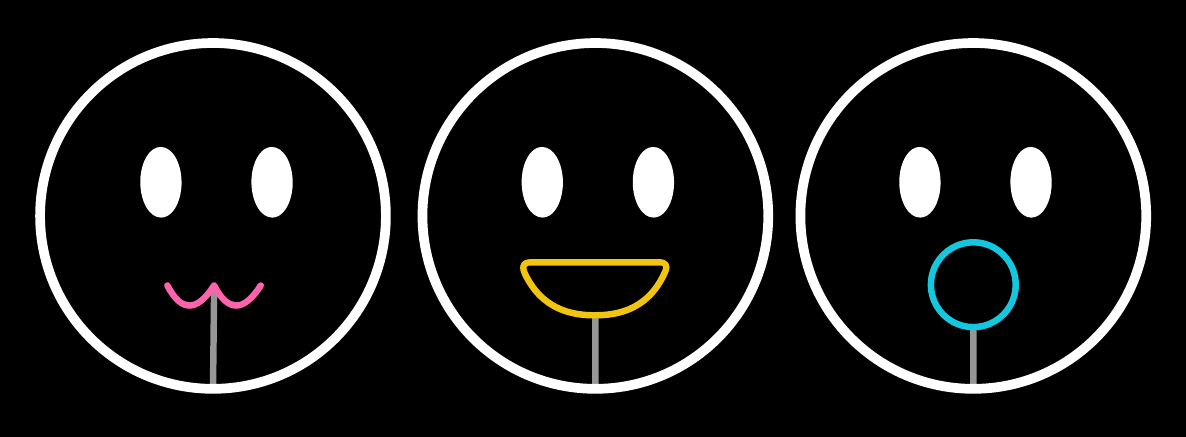} \hfill
        \includegraphics[height=0.212\textwidth]{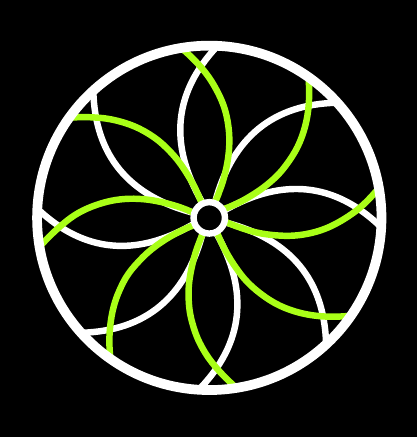} 
    \end{minipage}
    \caption{Potential applications of the Clo(o)k system. The programmable rings can carry elements beyond hour and minute hands: (1) Planetary Movement – independently rotating rings carry planets (Earth and Saturn) around a central Sun. (2) Interactive Face – a fixed front ring with eyes remains static, while the back ring with various attached mouth shapes rotates to show different facial expressions based on camera input. (3) Kinetic Art – two rings with dynamic patterns rotate at varying speeds and directions, creating mesmerizing visuals.}
    \Description{The first subfigure on the left illustrates a planetary motion simulation, where two small circles, representing planets, orbit a central sun on independent rotating rings. The planets follow separate paths, emphasizing their distinct orbits. The second subfigure in the center depicts an interactive facial expression application, showing three variations of a simple face with fixed eyes and different mouth shapes—a wavy line, a neutral curve, and a circular shape—suggesting dynamic expression changes as the back ring rotates. The third subfigure on the right presents an artistic pattern with multiple interwoven curved lines forming a symmetrical floral-like structure. These lines appear to be dynamically rotating, creating a mesmerizing visual effect through overlapping loops.}
    \label{fig:future}
\end{figure}

We designed \cl{}, a clock system that enables playful human-time interactions, suggesting a novel use of tangible, everyday objects to engage with intangible concepts. The demonstrated interactions represent only a subset of its potential uses. The detachable rings and programmable motors enable customization (Figure~\ref{fig:future}), inviting further exploration within and beyond the HCI community.

\begin{acks}
We thank Dr. Marcelo Coelho, Bill McKenna, Dominic Lim Co, and the students of MIT 4.031 for their valuable feedback on the design and fabrication of the \cl{}. We are grateful to Prof. Neil Gershenfeld, the TAs, and students of MAS.863 for their guidance and support on the electronics. We appreciate Prof. Hiroshi Ishii and his Tangible Interfaces class for inspiring the interaction design. Finally, we extend our thanks to everyone who visited our booth at the Open House of MAS.863 for their insightful comments and suggestions.
\end{acks}

\bibliographystyle{ACM-Reference-Format}
\bibliography{clook-new}

\end{document}